\documentclass[11pt]{amsart}

\usepackage{amssymb}
\usepackage{epsfig}
\usepackage{comment}
\usepackage{esint}
\usepackage{natbib}
\usepackage{marginnote}

\theoremstyle{definition}

\theoremstyle{remark}

\begin{document}
\title{Data-Driven Problems in Elasticity}

\title{Functional optimality of the sulcus pattern of the human brain}


\author[S.~Heyden and M.~Ortiz]
{S.~Heyden$^1$ and M.~Ortiz$^{2}$}

\address
{
    $^1$ Hausdorff Center for Mathematics,
    Endenicher Allee 60,
    53115 Bonn, Germany.
}

\address
{
    $^2$ Division of Engineering and Applied Science,
    California Institute of Technology,
    1200 E.~California Blvd., Pasadena, CA 91125, USA.
}

%
%
%
%
%
%

\begin{abstract}
We develop a mathematical model of information transmission across the biological neural network of the human brain. The overall function of the brain consists of the emergent processes resulting from the spread of information through the neural network. The capacity of the brain is therefore related to the rate at which it can transmit information through the neural network. The particular transmission model under consideration allows for information to be transmitted along multiple paths between points of the cortex. The resulting transmission rates are governed by potential theory. According to this theory, the brain has preferred and quantized transmission modes that correspond to eigenfunctions of the classical Steklov eigenvalue problem, with the reciprocal eigenvalues quantifying the corresponding transmission rates. We take the model as a basis for testing the hypothesis that the sulcus pattern of the human brain has evolved to maximize the rate of transmission of information between points in the cerebral cortex. We show that the introduction of sulci, or cuts, in an otherwise smooth domain indeed increases the overall transmission rate. We demonstrate this result by means of numerical experiments concerned with a spherical domain with a varying number of slits on its surface.
\end{abstract}

\maketitle

\section{Introduction}

The complex convoluted structure of the cerebral cortex is a hallmark of many mammalian brains and, in particular, of the human brain. Its specific surface morphology and the associated mechanisms underlying its growth
remain the subject of active topical research. Open challenges range from identifying the triggers for cortical folding processes to understanding the correlation between brain structure and brain function.

Mammalian brains are composed of an outer cortical layer, which is referred to as the {\sl grey matter} and which primarily consists of neuronal and glial cell bodies. In contrast, the inner subcortical core acts as a transmission structure and comprises axons connecting the neurons on the cerebral cortex. Such axons are coated with the electrically insulating substance myelin, which is white in appearance and lends the subcortical core the name {\sl white matter}. Axons typically show complex tree structures with myriads of branch points, and they make connections with other cells at junctions called {\sl synapses}. The profuse branching of axons results in a total of approximately $0.15$ quadrillion synapses, an exceedingly large number when compared to an estimated $86$ billion neurons in the average adult male human brain \citep{Hormuzdi:2004}. Due to the geometrical constraint imposed by the cranium, within which the brain must be embedded, the formation of cortical folds is the only way of increasing the cerebral surface area, which in turn results in an increase in the total number of neurons and a simultaneous minimization of their relative separation distance~\citep{Zilles:2013}. Since the total number of neurons and axons as well as the length of connections directly correlates to signaling speed and total information transmission rate, the cerebral surface area is viewed as a strong indicator of intelligence~\citep{Roth:2005}. Furthermore, several studies spanning diverse neurological disorders have revealed that the degree of cortical folding can serve as an indicator of abberations in brain development. For instance, reduced gyrification is found to be an inherent feature of the brain pathology in schizophrenia \citep{Nesvag:2014}, whereas increased frontal cortical folding is observed in the context of autism \citep{Hardan:2004}.

The particular mechanisms that drive the folding process during growth are as yet not fully understood. Van Essen~\citep{VanEssen:1997} first introduced the axonal tension hypothesis which assumes that tension along axons in the human connectome is the primary driving force for cortical folding. Based on this theory, axonal tension acts against hydrostatic pressure that is generated internally by the cerebrospinal fluid. Since long-distance connections enter and leave the cortex exclusively through the subcortical white matter, strongly interconnected regions will be pulled together during brain development due to tension along axons, whereas weakly connected regions are allowed to drift apart. Neighboring areas that are weakly interconnected are thus separated by outward folds, or {\sl gyri}, whereas strongly interconnected areas are separated by inward folds, or {\sl sulci}. As an alternative theory, the grey matter hypothesis postulates growth processes during cortical development to be governed by the driving forces of differential growth~\citep{Richman:1975}. Based on this model, folding processes are predicted when the growth of the outer (supragranular) cortical layer exceeds that of the inner (infragranular) layer~\citep{Zilles:2013}. Differential growth is thus introduced into the theory as a mechanism to release residual mechanical stresses by allowing for surface buckling. More recent studies combine the hypotheses of tension-mediated and differential-growth-induced cortical folding by modeling the cortex as a morphogenetically growing outer layer in combination with the subcortex as a strain-driven growing inner core~\citep{Budday:2014}. Both theories of differential growth and axonal tension, however, disagree with experimental findings. While dissection experiments did not reveal significant tangential tension within developing gyri as postulated by the axonal tension hypothesis~\citep{Xu:2010}, the differential growth hypothesis relies on unrealistic differences in stiffness parameters between the cortex and the underlying subcortical layers~\citep{Bayly:2014}.

Rather than {\sl how} the human brain cortex folds during growth, in this work we address the question of {\sl why} it is folded in the first place. We suppose that the overall function of the brain consists of the emergent processes resulting from the spread of information through the neural network. The capacity of the brain is therefore related to the rate at which it can transmit information through the neural network. The hypothesis underlying the present work is that the sulcus pattern of the human brain has evolved to maximize the rate of transmission of information between points in the cerebral cortex.

In order to render the problem in tractable mathematical terms, we formulate a simple graph model that relates the rate of transmission between points of the cortex to a problem in potential theory (cf.~\citep{Avena:2014, Misic:2015} for the biological basis of this model) and the classical Steklov eigenvalue problem \citep{Steklov:1902}. The particular transmission model under consideration allows for information to be transmitted along multiple paths between points of the cortex. We regard the human brain as a {\sl biological neural network}, i.~e., a collection of interconnected neurons. The interface between neurons consists of several axon terminals connected via synapses to dendrites. At any time, a neuron in the network can have an activation in the form of an action potential spike. The activation then spreads to all other connected neurons, which in turn become activated. From an information-theoretical point of view, each neuronal activation may be regarded as a bit of information. The overall function of the brain then consists of the emergent processes resulting from the spread of information through the neural network. The capacity of the brain is therefore related to the rate at which it can transmit information through the neural network.

A remarkable outcome of this model is that the transmission of information within the brain is {\sl quantized}. Thus, we find that the brain has preferred transmission modes that correspond to eigenfunctions of the classical Steklov eigenvalue problem \citep{Steklov:1902}, with the reciprocal eigenvalues quantifying the corresponding transmission rates. The Steklov spectrum of the brain thus collects all the preferred modes of transmission, or {\sl eigenfunctions}, of the brain. The Steklov eigenvalue problem originally arose in connection with hydrodynamics and has been extensively studied (cf.~\citep{Girouard:2017} for a recent review). The Steklov spectrum is discrete and the reciprocal Steklov eigenvalues, or transmission rates, have an accumulation point at zero \citep{Moiseev:1964, Kopachevskii:2001, Brock:2001}.

The model may be taken as a basis for testing the hypothesis that the sulcus pattern of the human brain has evolved to maximize the rate of transmission of information between points on the cerebral cortex. Specifically, the question is whether the sulcus pattern may be understood as the shape that minimizes the Steklov eigenvalues among all domains contained within a fixed confining set (the cranium). This type of shape optimization differs somewhat from the classical Steklov shape optimization problem which is concerned with competitor domains of fixed measure (cf., e.~g., \citep{Bogosel:2017, Girouard:2017}). While the full shape optimization problem is beyond the scope of this paper, we take some steps in that direction. In particular, we show that the introduction of sulci, or cuts, in an otherwise smooth domain indeed increases the overall transmission rate. We additionally demonstrate this result by means of numerical experiments concerned with spherical domains with a varying number of slits on its surface.

\section{Formulation of the problem}

In order to formulate a mathematical model of information transmission in the brain, we consider a simple neural network in the form of a cubic lattice.\footnote{We follow the notation of \citep{Treves:1970}, which may be consulted for background on the connection between Brownian motion and potential theory.} The nodes of the lattice represent the neurons and the bonds the synapses. Other lattices, including random networks, can be treated likewise without essential change in the outcome. Let $a \mathbb{Z}^3$ denote the lattice of points $x = (a l_1, a l_2, a l_3)$, with $a > 0$ the lattice parameter and $l_1$, $l_2$, $l_3 \in \mathbb{Z}$, where here and subsequently $\mathbb{Z}$ denotes the set of integer numbers. Two points $x$, $y \in a \mathbb{Z}^3$ are nearest neighbors if $|x-y| = a$. A path in $a \mathbb{Z}^3$ is a sequence of points such that every consecutive pair is also a pair of neighboring points.

Next, we turn to the propagation of signals through the lattice. Consider a signal that starts at $x$ and subsequently traverses a path in the lattice. At every point along the path, the signal has the choice of moving to one of the $6$ neighbors of the point with probability $1/6$ (cf.~\citep{Avena:2014, Misic:2015} for the biological basis of this model). By these set of rules, the paths available to the lattice for the transmission of information are {\sl random walks}. If we define the one-step transition probability as
\begin{equation}\label{w2leCr}
    p(x,y)
    =
    \left\{
    \begin{array}{ll}
        1/6 , & \text{if } x,\ y\ \text{neighbors}, \\
        0 , & \text{otherwise} .
    \end{array}
    \right.
\end{equation}
Then, by the Markov property of random walks,
\begin{equation}\label{Tr5emi}
    p(\Gamma) = p(x,z_1) p(z_1,z_2) \cdots p(z_{k-1},y)
\end{equation}
is the probability that the signal traverse a path $\Gamma = \{x, z_1, z_2, \dots, z_{k-1}, y\}$ joining $x$ to $y$. A frequentist interpretation of these probabilities is that signals have a choice of paths to travel between points of the neural lattice and that (\ref{Tr5emi}) gives the frequency with which a signal originating at $x$ traverses a particular path $\Gamma$ to reach another point $y$.

There is a well-known connection between random walks and harmonic functions (cf., e.~g., \citep{Treves:1970}). We recall that a function $u : a\mathbb{Z}^3 \to \mathbb{R}$ is {\sl harmonic} if
\begin{equation}
    u(x) = \frac{1}{6} \sum_{|x-y|=a} u(y) ,
\end{equation}
i.~e., if its value at every point $x$ of the lattice equals its average over the neighbors of $x$. The one-step shift or averaging operator $u \to P u$ is defined as
\begin{equation}\label{d1aplA}
    P u(x) = \sum_{y \in a \mathbb{Z}^3} p(x,y) \, u(y) ,
\end{equation}
and the discrete Laplacian as
\begin{equation}\label{j9UDri}
    \Delta = - 2 ( I - P ) ,
\end{equation}
where $I$ is the identity. From these definitions, it follows that a function $u$ is harmonic if and only if
\begin{equation}
    \Delta u = 0 ,
\end{equation}
i.~e., if it is a solution of Laplace's equation. By the maximum principle of harmonic functions, it follows that any bounded harmonic function over the entire lattice $a\mathbb{Z}^3$ is necessarily constant. Consider now the discrete Poisson equation
\begin{equation}\label{no0Sla}
    \Delta u + f = 0 ,
\end{equation}
where $f$ is a distribution of sources over $a \mathbb{Z}^3$. Since, as already noted, the kernel of the discrete Laplacian consists of constant functions, for solutions to exist the Fredholm alternative requires $f$ to sum to zero, in which case solutions are determined up to an additive constant. Inserting (\ref{j9UDri}) into (\ref{no0Sla}) and solving for $u$ we obtain
\begin{equation}
    u = \frac{1}{2} (I - P)^{-1} f = G f ,
\end{equation}
where
\begin{equation}\label{S1iutO}
    G = \frac{1}{2} (I - P)^{-1}
\end{equation}
is the free-space discrete Green's function. By translation-invariance, we have $G(x,y) = F(x-y)$, where $F$ is the fundamental solution of the discrete Laplacian.

The sought connection between potential theory and transmission of information along lattice paths can now be forged as follows. Expanding (\ref{S1iutO}) in Neumann series gives
\begin{equation}
    G
    =
    \frac{1}{2}
    \sum_{k=0}^\infty P^k .
\end{equation}
With $P$ as in (\ref{d1aplA}), it is readily shown that this series converges uniformly. Evaluating the powers $P^k$ explicitly, we obtain
\begin{equation}\label{xoA2ri}
\begin{split}
    G(x,y)
    & =
    \frac{1}{2}
    \Big(
        \delta(x-y)
        +
        P(x,y)
        \\ & +
        \sum_{k=2}^\infty
        \sum_{z_1\in a\mathbb{Z}^3}
        \cdots
        \sum_{z_{k-1}\in a\mathbb{Z}^3}
            p(x,z_1) p(z_1,z_2) \cdot p(z_{k-1},y)
    \Big) ,
\end{split}
\end{equation}
where
\begin{equation}
    \delta(x)
    =
    \left\{
    \begin{array}{ll}
        1 , & \text{if } x = 0, \\
        0 , & \text{otherwise} ,
    \end{array}
    \right.
\end{equation}
is the discrete Dirac function. For $x\neq y$, it follows from (\ref{w2leCr}) and (\ref{d1aplA}) that the term $P(x,y)$ in (\ref{xoA2ri}) contributes to the sum only if $y$ is a neighbor of $x$. Likewise, the product $p(x,z_1) p(z_1,z_2) \cdot p(z_{k-1},y)$ is non-zero only if the points $\{x, z_1, z_2, \dots, z_{k-1}, y\}$ define a path $\Gamma$ joining $x$ to $y$. Thus, eq.~(\ref{xoA2ri}) can be recast in the revealing form
\begin{equation}\label{zL1pro}
    G(x,y)
    =
    \frac{1}{2}
    \Big(
        \delta(x-y)
        +
        \sum_{\Gamma\in\mathcal{P}(x,y)} p(\Gamma)
    \Big) ,
\end{equation}
where $p(\Gamma)$ is defined in (\ref{Tr5emi}) and $\mathcal{P}(x,y)$ denotes the set of all paths joining $x$ to $y$ in the lattice. The path-sum representation (\ref{zL1pro}) shows that the Green's function $G(x,y)$ of the discrete Laplacian is the sum of contributions $p(\Gamma)$ arising from all paths joining $x$ to $y$.

In the neural network representation of the brain, we may regard $p(\Gamma)$, eq.~(\ref{Tr5emi}), as the probability that a signal originating at $x$ reach $y$ through the path $\Gamma\in\mathcal{P}(x,y)$. From representation (\ref{zL1pro}), it then follows that $G(x,y)$ is the total rate at which information injected into the network at $x$ is transmitted to $y$. This identification establishes the sought connection between the transmission of information through the neural network and discrete potential theory. Since the size of the neurons is much smaller than the overall size of the brain, we may further expect the continuum limit $a\to 0$ to supply a good approximation. This continuum limit can indeed be effected rigorously \citep{Treves:1970}. However the analysis is technical and beyond the scope of the present work. Therefore, in the sequel we proceed formally and simply replace the preceding discrete framework by its formally equivalent continuum counterpart.

\section{Connection with the Steklov eigenvalue problem}
\label{pHiad7}

A connection between the preceding graph model and the classical Steklov eigenvalue problem can be forged as follows. We consider a neural network occupying a domain $\Omega$. We further consider a distribution $h \in H^{-1/2}(\partial\Omega)$ of signals exchanged between points of the cortex $\partial\Omega$, with
\begin{equation}\label{S6awri}
    \int_{\partial\Omega} h(x) \, dS(x) = 0 .
\end{equation}
In the continuum limit, the corresponding transmission rate is
\begin{equation}
    R(h)
    =
    \int_{\partial\Omega}\int_{\partial\Omega}
        \frac{1}{2} G(x,y) h(x) h(y)
    \, dS(x) \, dS(y) ,
\end{equation}
or
\begin{equation}
    R(h)
    =
    \sup_{u \in H^1(\Omega)}
    \Big(
        \int_{\partial\Omega} h(x) u(x) \, dS(x)
        -
        \int_\Omega \frac{1}{2} |\nabla u(x) |^2 \, dx
    \Big) .
\end{equation}
We suppose that the brain has {\sl preferred transmission modes} that maximize $R(h)$ locally subject to the zero-sum condition (\ref{S6awri}) and the normalization constraint
\begin{equation}\label{RiUst0}
    \| h \|_{L^2(\partial\Omega)}^2
    =
    \int_{\partial\Omega} h^2(x) \, dS(x)
    =
    1 .
\end{equation}
We can combine the maximum transmission rate objective and the constraints into the Lagrangian
\begin{equation}\label{7hIeso}
\begin{split}
    F(u,h)
    & =
    \int_{\partial\Omega} h(x) u(x) \, dS(x)
    -
    \int_\Omega \frac{1}{2} |\nabla u(x) |^2 \, dx
    \\ & -
    \lambda
    \Big(
        \int_{\partial\Omega} h^2(x) \, dS(x) - 1
    \Big)
    -
    \mu
    \int_{\partial\Omega} h(x) \, dS(x) ,
\end{split}
\end{equation}
where $\lambda$ and $\mu$ are Lagrange multipliers, to be maximized with respect to $u \in H^1(\Omega)$ and $h \in H^{-1/2}(\partial\Omega)$. The stationarity of $F$ demands that
\begin{equation}\label{bri0Co}
    u(x) - 2 \lambda h(x) - \mu = 0 .
\end{equation}
From the zero-sum condition (\ref{S6awri}), we find
\begin{equation}
    \mu
    =
    \frac{1}{|\partial\Omega|}
    \int_{\partial\Omega} u(x) \, dS(x) ,
\end{equation}
where $|\partial\Omega|$ denotes the area of $\partial\Omega$. Inserting into (\ref{bri0Co}) and solving for $h(x)$, we obtain
\begin{equation}
    h(x)
    =
    \frac{1}{2\lambda}
    \Big(
        u(x)
        -
        \frac{1}{|\partial\Omega|}
        \int_{\partial\Omega} u(y) \, dS(y)
    \Big) .
\end{equation}
We thus conclude that the transmission mode $h(x)$ admits the representation
\begin{equation}\label{zoaB3a}
    h(x)
    =
    \frac{1}{2\lambda} v(x) ,
\end{equation}
with
\begin{equation}\label{froEs8}
    \int_{\partial\Omega} v(x) \, dS(x) = 0 .
\end{equation}
From the normalization constraint (\ref{RiUst0}), we additionally find
\begin{equation}
    2\lambda = \| v \|_{L^2(\partial\Omega)} ,
\end{equation}
whereupon (\ref{zoaB3a}) becomes
\begin{equation}\label{b6lAyi}
    h(x)
    =
    \frac{v(x)}{\| v \|_{L^2(\partial\Omega)}} .
\end{equation}
This expression yields a general representation of the transmission modes $h(x)$ in terms of potentials $v(x)$ with zero mean trace (\ref{froEs8}) over the boundary. Inserting (\ref{b6lAyi}) into (\ref{7hIeso}) and using (\ref{froEs8}), we obtain the reduced functional
\begin{equation}
    F(v)
    =
    \frac{1}{2\| v \|_{L^2(\partial\Omega)}}
    \int_{\partial\Omega} v^2(x) \, dS(x)
    -
    \int_\Omega \frac{1}{2} |\nabla v(x) |^2 \, dx ,
\end{equation}
to be maximized locally with respect to $v \in H^1(\Omega)$ subject to the zero mean trace constraint (\ref{froEs8}). The corresponding Euler-Lagrange equations are
\begin{subequations}\label{Choum4}
\begin{align}
    &
    \Delta v = 0 ,
    &
    \text{in } \Omega ,
    \\ &
    \frac{\partial v}{\partial {n}}
    =
    \sigma \, v ,
    &
    \text{on } \partial\Omega ,
\end{align}
\end{subequations}
where ${n}$ is the outward unit normal and we write
\begin{equation}
    \sigma
    =
    \frac{1}{\| v \|_{L^2(\partial\Omega)}} .
\end{equation}
We recognize (\ref{Choum4}) as the classical {\sl Steklov eigenvalue problem} \citep{Steklov:1902}.

It is well-known that the Steklov eigenvalue problem has a {\sl discrete spectrum} (cf., e.~g., \citep{Girouard:2017})
\begin{equation}\label{3IucHo}
    0 < \sigma_1 \leq \sigma_2 \leq \sigma_3 \leq \dots \to +\infty
\end{equation}
as long as the trace operator $H^1(\Omega) \to L^2(\partial\Omega)$ is compact \citep{Arendt:2012}. This property holds under suitable regularity of the domain, e.~g., if $\Omega$ has Lipschitz boundary \citep{Necas:2012}.

\section{Monotonicity with respect to cutting}
\label{ChI7rl}

The Steklov eigenvalues have a number of domain monotonicity properties that bear directly on the present discussion (cf., e.~g., \cite{Kulczycki:2009}). We recall that the Steklov eigenvalues admit the variational characterization (cf., e.~g., \cite{Lamberti:2015, Bogosel:2017, Girouard:2017})
\begin{equation}\label{S3oeTi}
    \sigma_n
    =
    \inf_{V_n \subset H^1(\Omega)/\mathbb{R}}
    \ \sup_{v \in V_n \backslash \{0\}}
    \frac
    {
        \int_\Omega |\nabla v(x) |^2 \, dx
    }
    {
        \int_{\partial\Omega} v^2(x) \, dS(x)
    } ,
\end{equation}
where the infimum is taken over all $n$-dimensional subspaces $V_n$ of $H^1(\Omega)/\mathbb{R} = \{ u \in H^1(\Omega),\ \int_{\partial\Omega} u(x) \, dS(x) = 0\}$. The corresponding eigenfunctions $v_n$ represent {\sl transmission modes} of the brain and the inverse eigenvalues $1/\sigma_n$ give the corresponding {\sl transmission rates}.

The beneficial effect of sulci to brain function can be deduced from (\ref{S3oeTi}) as follows. Let $\Omega$ be the domain of the cranium and let $\{\sigma_n\}$ be its Steklov spectrum. Let $\Gamma$ be a collection of cuts (sulci) performed on the boundary $\Omega$, set $\Omega'=\Omega \backslash \Gamma$, and let $\{\sigma_n'\}$ be the Steklov spectrum of the slit domain $\Omega'$. From the variational characterization (\ref{S3oeTi}) of the Steklov eigenvalues, we have
\begin{equation}
\begin{split}
    \sigma_n'
    & =
    \inf_{V'_n \subset H^1(\Omega')/\mathbb{R}}
    \ \sup_{v' \in V'_n \backslash \{0\}}
    \frac
    {
        \int_{\Omega'} |\nabla v'(x) |^2 \, dx
    }
    {
        \int_{\partial\Omega'} v'^2(x) \, dS(x)
    }
    \\ & =
    \inf_{V'_n \subset H^1(\Omega')/\mathbb{R}}
    \ \sup_{v' \in V'_n \backslash \{0\}}
    \frac
    {
        \int_{\Omega'} |\nabla v'(x) |^2 \, dx
    }
    {
        \int_{\partial\Omega} v'^2(x) \, dS(x)
        +
        \int_\Gamma v'^2(x) \, dS(x)
    }
    \\ & \leq
    \inf_{V'_n \subset H^1(\Omega')/\mathbb{R}}
    \ \sup_{v' \in V'_n \backslash \{0\}}
    \frac
    {
        \int_{\Omega'} |\nabla v'(x) |^2 \, dx
    }
    {
        \int_{\partial\Omega} v'^2(x) \, dS(x)
    }
    \\ & \leq
    \inf_{V_n \subset H^1(\Omega)/\mathbb{R}}
    \ \sup_{v \in V_n \backslash \{0\}}
    \frac
    {
        \int_\Omega |\nabla v(x) |^2 \, dx
    }
    {
        \int_{\partial\Omega} v^2(x) \, dS(x)
    }
    =
    \sigma_n ,
\end{split}
\label{eq:SigmaSlit}
\end{equation}
since $H^1(\Omega) \subset H^1(\Omega')$. We thus conclude that the introduction of {\sl sulci increases the transmission rate of all the transmission modes of the brain}. The same argument shows that the transmission rate also increases when existing cuts are made deeper. These monotonicity properties explain the functional benefit of sulci to the function of the brain.

\section{Numerical experiments}
\label{1iAstl}

We further illustrate the connection between sulci and rate of transmission by means of selected numerical experiments. Specifically, we compare the Steklov eigenvalues of a sphere with those of slit-spheroidal domains and show that the introduction of slits does indeed reduce the Steklov eigenvalues. We carry out all calculations be recourse to the finite element method. The accuracy of the method is assessed and controlled with the aid of the known Steklov spectrum of the spherical domain.

\subsection{Spherical domain}

In order to compute the Steklov spectrum it is convenient to reformulate the problem as an eigenvalue problem for the Laplace operator with Neumann boundary conditions,
\begin{subequations} \label{eq:LaplaceNeumann}
\begin{align}
    &
    \Delta v = \sigma \, \rho \, v ,
    &
    \text{in } \Omega ,
    \\ &
    \frac{\partial v}{\partial {n}}
    =
    0 ,
    &
    \text{on } \partial \Omega ,
\end{align}
\end{subequations}
with mass density $\rho$ concentrated on the boundary $\partial\Omega$. The eigenvalues of problem (\ref{eq:LaplaceNeumann}) are known to coincide with those of the classical Steklov eigenvalue problem \citep{Lamberti:2015,Arrieta:2008}. With $\Omega$ the unit ball in $\mathbb{R}^n$ and $n\geq 2$, the Steklov eigenvalues are given explicitly by the sequence
\begin{equation}
	\sigma_j = j, \quad j\in\mathbb{N},
\end{equation}
and the corresponding eigenfunctions are the homogeneous harmonic polynomials of degree $j$ \citep{Lamberti:2017}. Furthermore, the eigenvalues have multiplicity
\begin{equation}
	\text{Mult}(\sigma_j) = \begin{pmatrix} N+j-1 \\ N-1 \end{pmatrix} - \begin{pmatrix} N+j-3 \\ N-1 \end{pmatrix}.
\end{equation}

\begin{table}
\begin{tabular}{l*{9}{c}r}
					 & $\sigma_0$ & $\sigma_1$ & $\sigma_2$ & $\sigma_3$ & $\sigma_4$  & $\sigma_5$ & $\sigma_6$ & $\sigma_7$  & $\sigma_8$ \\
\hline
Analytical & 0 & 1 & 1 & 1 & 2 & 2 & 2 & 2 & 2 \\
Numerical  & 0 & 0.997 & 0.997 & 0.997 & 2.00 & 2.00 & 2.00 & 2.00 & 2.00  \\
\end{tabular}
\caption{Lowest nine eigenvalues of the computed Steklov spectrum of the unit ball in comparison to analytical results.}
\label{tab:Steklov_Analytical}
\end{table}

We assess the accuracy of the finite element discretization by means of the unit ball test case just described.We discretize $\Omega$ by means of $\approx 11,000$ linear tetrahedral elements and tile the boundary $\Omega$ by means of $\approx 2,000$ membrane elements of uniform areal mass density. The interior degrees of freedom are subsequently eliminated by means of static condensation, resulting in an eigenvalue problem involving the surface degrees of freedom only.

\begin{figure}[h!]
\begin{center}
    \includegraphics[width=0.95\linewidth]{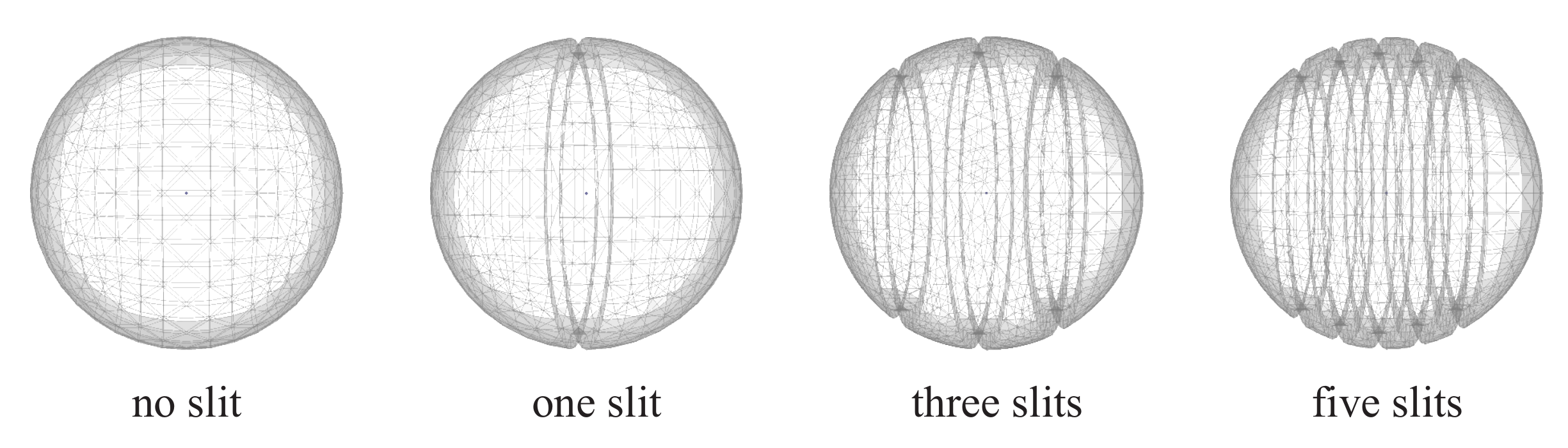}
    \caption{Finite element meshes for different slit-spheroidal domains (shown for a translucent solid) with an increasing number of slits.}
\end{center}
\label{fig:Meshes}\vspace*{-9pt}
\end{figure}

The $9$ lowest eigenvalues of the computed Steklov spectrum are compared in Table~\ref{tab:Steklov_Analytical} to the exact analytical values. As can be seen from the table, the lowest non-zero eigenvalue is computed with a relative error of $0.3\,\%$. In general, the numerical accuracy of eigenvalues $\sigma_j$ is known to decrease with increasing $j$. For instance the eigenvalue $\sigma_{25}=5$ is obtained with a relative error of $1.0\,\%$. We further note that numerical eigenvalues have the expected multiplicities.

\begin{figure}[h]
\begin{center}
    \includegraphics[width=0.95\linewidth]{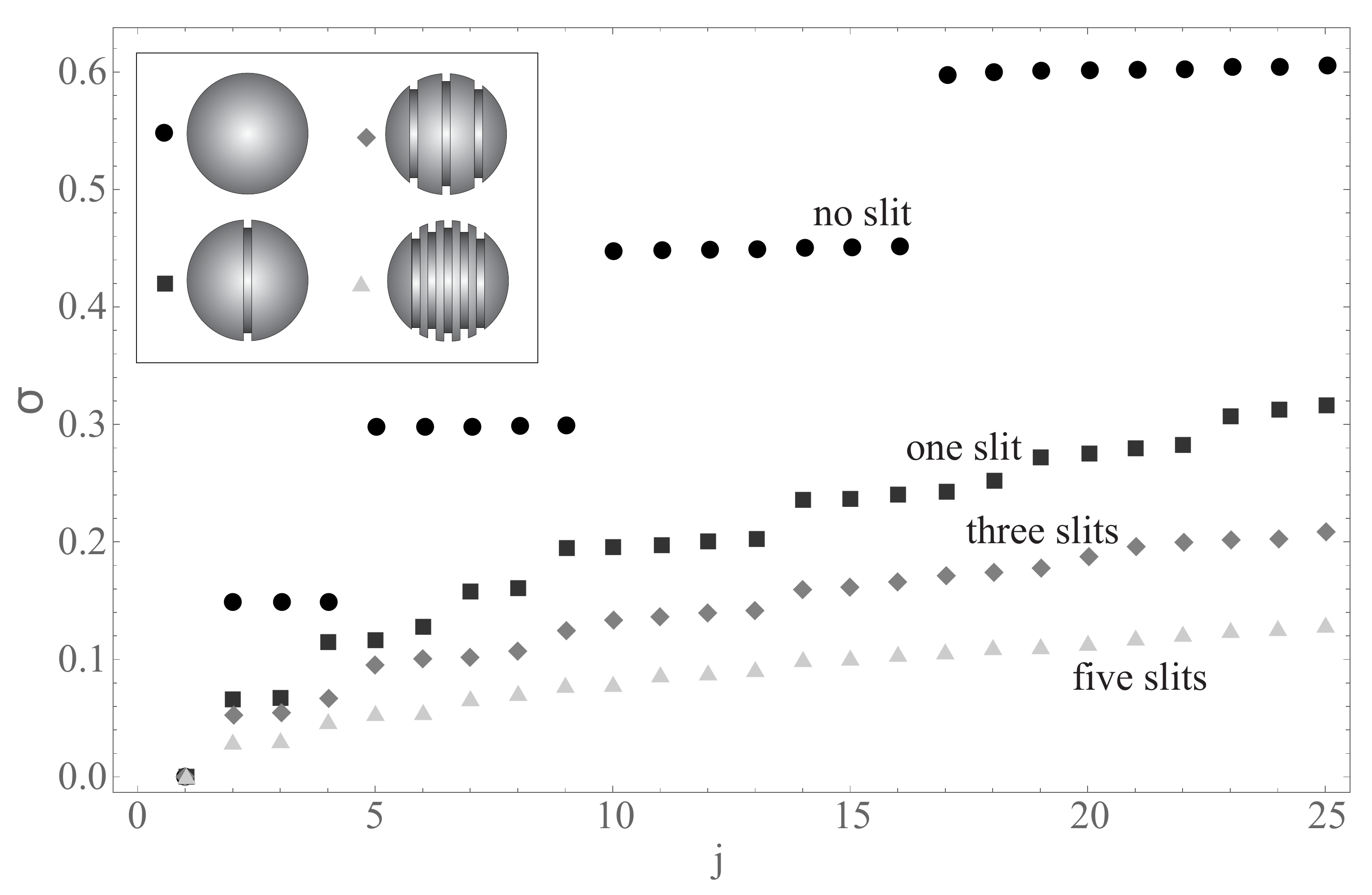}
    \caption{Steklov spectra of the lowest $25$ eigenvalues $\sigma_j$ of slit-spheroidal geometries for an increasing number of slits.}
\end{center}
\label{fig:SteklovSlitSpheroids}\vspace*{-9pt}
\end{figure}

\subsection{Slit-spheroidal domains}

Next we proceed to compute the Steklov spectrum of slit-spheroidal domains. To this end, we start with a spherical domain of radius $6.7$\,cm, which matches the average male human brain volume of $1260$\,cm$^{3}$ \citep{Cosgrove:2007}. Subsequently, slit-spheroidal domains are constructed by successively introducing slits of varying depth, intended to represent idealized sulci. Geometry definition as well as meshing are performed using the Siemens PLM Software Femap with NX Nastran. Representative geometries and finite element meshes are shown in Fig.~1.

Fig.~2 shows the lowest $25$ Steklov eigenvalues $\sigma_j$ of slit-spheroidal geometries with up to $5$ slits. The finite element meshes used in calculations contain $\approx 7,000$ linear tetrahedral elements and $\approx 5,000$ membrane elements. As may be seen from the figure, the eigenvalues are found to decrease for an increasing number of slits.

To further elucidate the influence of the slit depth, Fig.~3 shows the Steklov spectra of the lowest $25$ eigenvalues $\sigma_j$ of a one-slit-spheroidal geometry for an increasing depth of the slit while keeping the width of the slit constant. Meshes with a slit depth of $0.67$\,cm as well as $2.68$\,cm are generated and the calculated eigenvalues are depicted in comparison to corresponding values for the smooth domain. As may be seen from the figure, all eigenvalues decrease with increasing slit depth.

These results are in keeping with the monotonicity arguments of Section~\ref{ChI7rl}, which show that the Steklov eigenvalues (respec., transmission rates) decrease (respec., increase) when cuts are introduced in a domain and when the cuts are made deeper.

\begin{figure}[h]
\begin{center}
    \includegraphics[width=0.95\linewidth]{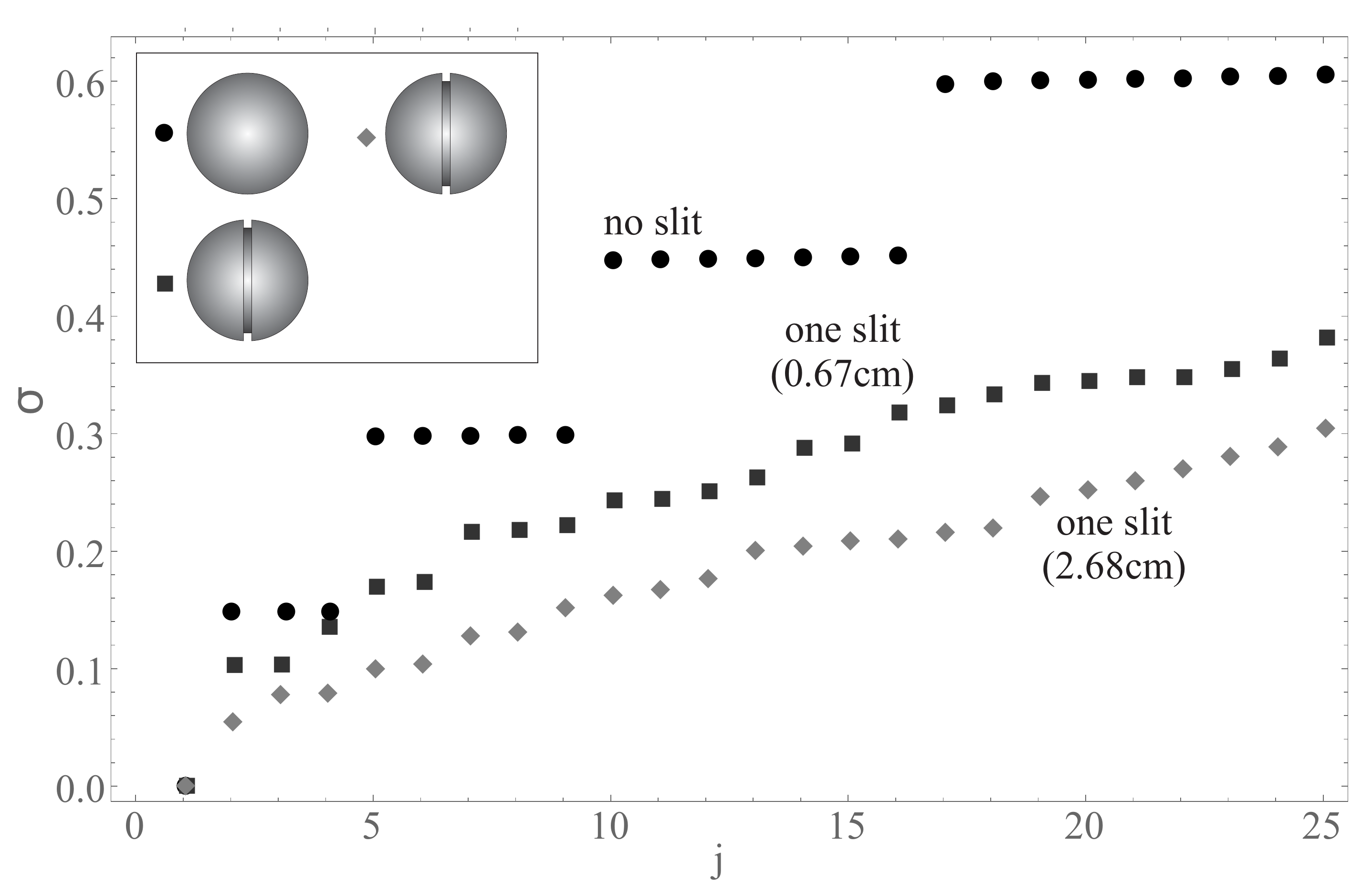}
    \caption{Steklov spectra of the lowest $25$ eigenvalues $\sigma_j$ of a one-slit-spheroidal geometry for an increasing depth of the slit.}
\end{center}
\label{fig:Slit_Depth}\vspace*{-9pt}
\end{figure}

\section{Concluding remarks}

This paper establishes a connection, through a neural network model, between information transmission between points in the brain cortex and the classical Steklov eigenvalue problem. The implications of this connection are manifold. Firstly, according to the model the modes of transmission of the brain are described spatially by the Steklov eigenfunctions and, therefore, are {\sl quantized}, with the corresponding eigenvalues supplying the transmission rates of the modes. In particular, the lowest Steklov eigenvalue gives the largest possible rate of transmission of the brain. However, the transmission rates $1/\sigma_j$ have an accumulation point at the origin, eq.~(\ref{3IucHo}), with the result that {\sl slow} transmission rates are densely distributed and quantization is ostensibly lost in that range. The physiological significance of the fast transmission modes, corresponding to the lowest Steklov eigenvalues, is yet to be elucidated.

We have also shown that the transmission efficiency of a domain increases when cuts are introduced. Specifically, the Steklov eigenvalues decrease upon the introduction of the cuts and decrease further as the cuts are made deeper. This type of monotonicity provides an explanation for the functional benefit of the {\sl sulci}, regarded as cuts in an otherwise smooth brain. However, since the transmission efficiency increases monotonically with the depth of the cuts, no regular brain shape can be expected to maximize transmission efficiency or, equivalently, minimize the Steklov eigenvalues. A well-posed optimization problem can be obtained by introducing additional constraints, e.~g., on the cortical surface area. Indeed, the classical shape optimization problems for the Steklov eigenvalues consider competing domains of fixed boundary area (cf., e.~g., \cite{Bogosel:2017, Girouard:2017}). Such a strong surface area constraint is not appropriate to the case at hand, in which the brain shapes of interest fill the cranium and differ only in the sulcus, or cut, pattern. An alternative is to penalize, instead of constraining, the cortical surface area. An objective function of this type is, for instance,
\begin{equation}\label{noe6lE}
    F(\Omega) = \sigma_1(\Omega) + A \, |\partial\Omega|^\alpha,
\end{equation}
for some constants $A$ and $\alpha$ and with $|\partial\Omega|$ the surface area. The additional term $A \, |\partial\Omega|^\alpha$ represents the physiological cost of cortical surface area. The problem is then to minimize $F$ over all brain domains $\Omega$ contained in the cranium $K$. Our conjecture is that the optimal shape is obtained by cutting $K$ into an optimal sulcus pattern, but a rigorous analysis of this conjecture is beyond the scope of the present work.

Objective functions of the form (\ref{noe6lE}) can also potentially explain scaling relations and size effects that have been uncovered through systematic analyses of the variation in cortical folding across large samples of mammalian species. For instance, \cite{Mota:2017} have shown that the degree of cortical folding scales uniformly across species, across individuals, and within individual cortices. Specifically, for all noncetacean gyrencephalic species they find the scaling relation
\begin{equation}\label{wrI4xo}
    |\partial\Omega| = \max \{ |\partial K|, B \, |\partial K|^\beta \},
\end{equation}
with $\beta \sim 1.242\pm 0.018$, $B \sim (1000 {\rm mm}^2)^{1-\beta}$. Remarkably, this power law is significantly superlinear, which implies that, as the total surface area increases, the brain becomes increasingly folded. We recall that the ratio $|\partial\Omega|/|\partial K|$ of total cortical area to exposed cortical area is related to the classical {\sl gyrification index} \citep{Zilles:1988}. The scaling law (\ref{wrI4xo}) shows that the gyrification index is a function of brain size, with larger brains having a more folded cortex.

Scaling relations such as (\ref{wrI4xo}) can be rationalized within the present framework as follows. Suppose that the fundamental transmission rate $1/\sigma_1(\Omega)$ obeys an {\sl optimal scaling law} of the form
\begin{equation}\label{ch0uGo}
    c
    |\partial K|^\epsilon |\partial\Omega|^\delta
    \leq
    \frac{1}{\sigma_1(\Omega)}
    \leq
    C
    |\partial K|^\epsilon |\partial\Omega|^\delta ,
\end{equation}
for some constants $C > c > 0$ and exponents $\epsilon$ and $\delta$. We recall that scaling laws are said to be {\sl optimal} if they entail power-law lower and upper bounds with {\sl matching exponents}, in this case $\epsilon$ and $\delta$. Optimal scaling laws were developed in mathematics in connection with energy-minimizing branched microstructures \citep{KM92, KM94, CKO99, Conti00}. From dimensional considerations, we must have
\begin{equation}\label{siE7ie}
    \epsilon + \delta = \frac{1}{2} .
\end{equation}
Suppose, in addition, that the optimal brain shape satisfies equipartition between the two terms of the objective function (\ref{noe6lE}). Then, we have
\begin{equation}
    \sigma_1(\Omega) \propto  |\partial\Omega|^\alpha .
\end{equation}
Inserting into (\ref{ch0uGo}) and assuming that the upper and lower bounds are tight, we have
\begin{equation}
    |\partial\Omega| \propto |\partial K|^{\epsilon/(\alpha-\delta)} .
\end{equation}
Comparing with (\ref{wrI4xo}) we finally find
\begin{equation}\label{spi0pR}
    \beta = \frac{\epsilon}{\alpha-\delta} ,
\end{equation}
which gives the exponent $\beta$ in terms of the optimal-scaling exponents $\epsilon$ and $\delta$ and the cost exponent $\alpha$. If, for instance, we assume that the cortical surface area cost is proportional to the cortical surface areas, $\alpha=1$, then (\ref{spi0pR}) specializes to
\begin{equation}\label{spi0pR}
    \beta
    =
    \frac{\epsilon}{1/2-\epsilon}
    =
    \frac{1/2-\delta}{1-\delta} ,
\end{equation}
where we have used (\ref{siE7ie}). Identity (\ref{spi0pR}) gives the gyrification exponent $\beta$ in terms of the optimal scaling exponents for the fundamental Steklov eigenvalue.

Beyond the question of gyrification, we remark that the Steklov eigenfunctions provide a convenient orthogonal basis for the spatial representation of brain activity. Thus, for a given patient-specific brain geometry $\Omega$ the Steklov eigenfunctions and eigenvalues can be computed numerically, e.~g., by recourse to the finite-element method as in Section~\ref{1iAstl} or by other means \citep{Akhmetgaliyev:2017}. We also recall that, according to the theory put forth in Section~\ref{pHiad7}, the brain activity is described by a potential $u$ over $\Omega$. It thus follows that any pattern of brain activity $u$ can be decomposed into, and then represented as a sum of, Steklov modal components.

We also note that the gradient of the potential, $J = \nabla u$, measures the electrical current density associated with the activity of the brain. By Amp\`ere's law, $\nabla \times B = \mu_0 J$, this current density induces a magnetic field $B = \nabla \times A$, where $A$ is a vector potential satisfying the gage condition $\nabla\cdot A = 0$. This vector potential, and the corresponding magnetic field, are given by the Biot-Savart law and extend outside the brain and the cranium. Though weak, the exterior fields can be detected and measured, e.~g., with the aid of sensitive detectors known as {\sl superconducting quantum interference devices} (SQUIDs). Such measurements are known as {\sl magnetoencephalograms}, or MEGs (cf., e.~g., \cite{Cohen:2004}).

A compelling alternative consists of reversing the process and inducing activity patterns in the brain through the application of external magnetic fields, a process known as {\sl transcranial magnetic stimulation} (TMS) (cf., e.~g., \cite{George:2000}). During a TMS procedure, a magnetic field generator in the form of a coil is placed near the head of the patient. Evidence suggests that TMS is effective against neuropathic pain and treatment-resistant major depressive disorder. However, the magnetic field generators, or coils, used at present are not tailored to patient-specific geometries and are relatively delocalized. The detailed knowledge, through patient-specific calculations, of the Steklov spectra of an individual brain opens up the way for stimulating specific transmission modes, including the fundamental modes at which the brain performs at its greatest capacity. Such mode-specific stimulation could be achieved by externally applying to the cranium shaped electromagnetic fields corresponding to the fundamental Steklov modes. The therapeutic benefits of such tailored TMS procedures are yet to be ascertained.

\section*{Acknowledgements}

SH gratefully acknowledges support from the Alexander von Humboldt Stiftung through a Research Fellowship for Postdoctoral Researchers.


\bibliographystyle{elsarticle-harv}
{\small\bibliography{Biblio}}

\begin{thebibliography}{35}
\expandafter\ifx\csname natexlab\endcsname\relax\def\natexlab#1{#1}\fi
\expandafter\ifx\csname url\endcsname\relax
  \def\url#1{\texttt{#1}}\fi
\expandafter\ifx\csname urlprefix\endcsname\relax\def\urlprefix{URL }\fi

\bibitem[{Akhmetgaliyev et~al.(2017)Akhmetgaliyev, Kao, and
  Osting}]{Akhmetgaliyev:2017}
Akhmetgaliyev, E., Kao, C.~Y., Osting, B., 2017. Computational methods for
  extremal steklov problems. SIAM J. Control Optim 55~(2), 1226–1240.

\bibitem[{Arendt and Mazzeo(2012)}]{Arendt:2012}
Arendt, W., Mazzeo, R., 2012. Friedlander's eigenvalue inequalities and the
  dirichlet-to-neumann semigroup. Communications on Pure and Applied Analysis
  11~(6), 2201--2212.

\bibitem[{Arrieta et~al.(2008)Arrieta, Jiminez-Casas, and
  Rodriguez-Bernal}]{Arrieta:2008}
Arrieta, J.~M., Jiminez-Casas, A., Rodriguez-Bernal, A., 2008. Flux terms and
  robin boundary con- ditions as limit of reactions and potentials
  concentrating in the boundary. Rev. Mat. Iberoam. 24~(1), 183--211.

\bibitem[{Avena-Koenigsberger et~al.(2014)Avena-Koenigsberger, Goni, Betzel,
  van~den Heuvel, Griffa, Hagmann, Thiran, and Sporns}]{Avena:2014}
Avena-Koenigsberger, A., Goni, J., Betzel, R.~F., van~den Heuvel, M.~P.,
  Griffa, A., Hagmann, P., Thiran, J.~P., Sporns, O., 2014. Using pareto
  optimality to explore the topology and dynamics of the human connectome.
  Philosophical Transactions of the Royal Society B - Biological Sciences
  369~(1653), 20130530.

\bibitem[{Bayly et~al.(2014)Bayly, Okamoto, Xu, Shi, and Taber}]{Bayly:2014}
Bayly, P.~V., Okamoto, R., Xu, G., Shi, Y., Taber, L.~A., 2014. A cortical
  folding model incorporating stress-dependent growth explains gyral
  wavelengths and stress patterns in the developing brain. Phys. Biol. 10~(1),
  016005.

\bibitem[{Bogosel et~al.(2017)Bogosel, Bucur, and Giacomini}]{Bogosel:2017}
Bogosel, B., Bucur, D., Giacomini, A., 2017. Optimal shapes maximizing the
  steklov eigenvalues. Siam Journal on Mathematical Analysis 49~(2),
  1645--1680.

\bibitem[{Brock(2001)}]{Brock:2001}
Brock, F., 2001. An isoperimetric inequality for eigenvalues of the stekloff
  problem. Zeitschrift f\"ur Angewandte Mathematik und Mechanik 81~(1), 69--71.

\bibitem[{Budday et~al.(2014)Budday, Steinmann, and Kuhl}]{Budday:2014}
Budday, S., Steinmann, P., Kuhl, E., 2014. The role of mechanics during brain
  development. J. Mech. Phys. Solids 72, 75--92.

\bibitem[{Choksi et~al.(1999)Choksi, Kohn, and Otto}]{CKO99}
Choksi, R., Kohn, R.~V., Otto, F., 1999. Domain branching in uniaxial
  ferromagnets: a scaling law for the minimum energy. Comm. Math. Phys. 201,
  61--79.

\bibitem[{Cohen and Halgren(2004)}]{Cohen:2004}
Cohen, D., Halgren, E., 2004. Magnetoencephalography. In: Adelman, G., Smith,
  B. (Eds.), Encyclopedia of Neuroscience. Elsevier.

\bibitem[{Conti(2000)}]{Conti00}
Conti, S., 2000. Branched microstructures: scaling and asymptotic
  self-similarity. Comm. Pure Appl. Math. 53, 1448--1474.

\bibitem[{Cosgrove et~al.(2007)Cosgrove, Mazure, and Staley}]{Cosgrove:2007}
Cosgrove, K.~P., Mazure, C.~M., Staley, J.~K., 2007. Evolving knowledge of sex
  differences in brain structure, function, and chemistry. Biological
  Psychiatry 62~(8), 847--855.

\bibitem[{George and Belmaker(2000)}]{George:2000}
George, M.~S., Belmaker, R.~H., 2000. Transcranial Magnetic Stimulation in
  Neuropsychiatry. American Psychiatric Press.

\bibitem[{Girouard and Polterovich(2017)}]{Girouard:2017}
Girouard, A., Polterovich, I., 2017. Spectral geometry of the steklov problem
  (survey article). Journal of Spectral Theory 7~(2), 321--359.

\bibitem[{Hardan et~al.(2004)Hardan, Jou, Keshavan, Varma, and
  Minshew}]{Hardan:2004}
Hardan, A.~Y., Jou, R.~J., Keshavan, M.~S., Varma, R., Minshew, N.~J., 2004.
  Increased frontal cortical folding in autism: a preliminary mri study.
  Psychiatry Research 131~(3), 263--268.

\bibitem[{Hormuzdi et~al.(2004)Hormuzdi, Filippov, Mitropoulou, Monyer, and
  Bruzzone}]{Hormuzdi:2004}
Hormuzdi, S.~G., Filippov, M., Mitropoulou, G., Monyer, H., Bruzzone, R., 2004.
  Electrical synapses: a dynamic signaling system that shapes the activity of
  neural networks. Biochimica et Biophysica Acta 1662~(1-2), 113--137.

\bibitem[{Kohn and M\"uller(1992)}]{KM92}
Kohn, R.~V., M\"uller, S., 1992. Branching of twins near an
  austenite-twinned-martensite interface. Phil. Mag. A 66, 697--715.

\bibitem[{Kohn and M\"uller(1994)}]{KM94}
Kohn, R.~V., M\"uller, S., 1994. Surface energy and microstructure in coherent
  phase transitions. Comm. Pure Appl. Math. 47, 405--435.

\bibitem[{Kopachevskii and Krein(2001)}]{Kopachevskii:2001}
Kopachevskii, N.~D., Krein, S.~G., 2001. Operator approach to linear problems
  of hydrodynamics. Operator theory, advances and applications. Birkhauser
  Verlag, Basel; Boston.

\bibitem[{Kulczycki and Kuznetsov(2009)}]{Kulczycki:2009}
Kulczycki, T., Kuznetsov, N., 2009. 'high spots' theorems for sloshing
  problems. Bulletin of the London Mathematical Society 41, 495--505.

\bibitem[{Lamberti and Provenzano(2013)}]{Lamberti:2015}
Lamberti, P.~D., Provenzano, L., 2013. Viewing the steklov eigenvalues of the
  laplace operator as critical neumann eigenvalues. In: Current Trends in
  Analysis and Its Applications,. Birkh\"{a}user Basel, pp. 171--178.

\bibitem[{Lamberti and Provenzano(2017)}]{Lamberti:2017}
Lamberti, P.~D., Provenzano, L., 2017. Neumann to steklov eigenvalues:
  asymptotic and monotonicity results. Proceedings of the Royal Society of
  Edinburgh Section a-Mathematics 147~(2), 429--447.

\bibitem[{Misic et~al.(2015)Misic, Betzel, Nematzadeh, Goni, Griffa, Hagmann,
  Flammini, Ahn, and Sporns}]{Misic:2015}
Misic, B., Betzel, R.~F., Nematzadeh, A., Goni, J., Griffa, A., Hagmann, P.,
  Flammini, A., Ahn, Y.~Y., Sporns, O., 2015. Cooperative and competitive
  spreding dynamics of the human connectome. Neuron 86~(6), 1518--1529.

\bibitem[{Moiseev(1964)}]{Moiseev:1964}
Moiseev, N.~N., 1964. Introduction to the theory of oscillations of
  liquid-containing bodies. In: Advances in Applied Mechanics. Vol.~8. Academic
  Press, New York, pp. 233--289.

\bibitem[{Mota and Herculano-Houzel(2017)}]{Mota:2017}
Mota, B., Herculano-Houzel, S., 2017. Cortical folding scales universally with
  surface area and thickness, not number of neurons. Science, Research Reports
  349~(6243), 74--77.

\bibitem[{Necas et~al.(2012)Necas, Simader, and Necasova}]{Necas:2012}
Necas, J., Simader, C.~G., Necasova, S., 2012. Direct methods in the theory of
  elliptic equations. Springer monographs in mathematics. Springer, Heidelberg
  New York.

\bibitem[{Nesv{\aa}g et~al.(2014)Nesv{\aa}g, Schaer, Haukvik, Westlye, Rimol,
  Lange, Hartberg, Ottet, Melle, Andreassen, J\"{o}nsson, Agartz, and
  Eliez}]{Nesvag:2014}
Nesv{\aa}g, R., Schaer, M., Haukvik, U.~K., Westlye, L.~T., Rimol, L.~M.,
  Lange, E.~H., Hartberg, C.~B., Ottet, M.~C., Melle, I., Andreassen, O.~A.,
  J\"{o}nsson, E.~G., Agartz, I., Eliez, S., 2014. Reduced brain cortical
  folding in schizophrenia revealed in two independent samples. Schizophrenia
  Research 152~(2-3), 333--338.

\bibitem[{Richman et~al.(1975)Richman, Steward, Hutchinson, and
  Caviness}]{Richman:1975}
Richman, D.~P., Steward, R.~M., Hutchinson, J.~W., Caviness, V.~S., 1975.
  Mechnical model of brain convolutional development. Science 189, 18--21.

\bibitem[{Roth and Dicke(2005)}]{Roth:2005}
Roth, G., Dicke, U., 2005. Evolution of the brain and intelligence. Trends
  Cogn. Sci. 9, 250--257.

\bibitem[{Steklov(1902)}]{Steklov:1902}
Steklov, M.~W., 1902. Sur les probl\`{e}mes fondamentaux de la physique
  math\'{e}matique. Ann. Sci. Ecole Norm. Sup. 19, 455--490.

\bibitem[{Treves(1970)}]{Treves:1970}
Treves, F., 1970. Linear partial differential equations. Notes on mathematics
  and its applications. Gordon and Breach, New York,.

\bibitem[{Van~Essen(1997)}]{VanEssen:1997}
Van~Essen, D.~C., 1997. A tension-based theory of morphogenesis and compact
  wiring in the central nervous system. Nature 385~(23), 313--318.

\bibitem[{Xu et~al.(2010)Xu, Knutsen, Dikranian, Kroenke, Bayly, and
  Taber}]{Xu:2010}
Xu, G., Knutsen, A.~K., Dikranian, K., Kroenke, C.~D., Bayly, P.~V., Taber,
  L.~A., 2010. Axons pull on the brain, but tension does not drive cortical
  folding. J. Biomech. Eng. 132~(7), 071013.

\bibitem[{Zilles et~al.(1988)Zilles, Armstrong, Schleicher, and
  Kretschmann}]{Zilles:1988}
Zilles, K., Armstrong, E., Schleicher, A., Kretschmann, H.~J., 1988. The human
  pattern of gyrification in the cerebral cortex. Anatomy and Embryology 179,
  173--179.

\bibitem[{Zilles et~al.(2013)Zilles, Palomero-Gallagher, and
  Amunts}]{Zilles:2013}
Zilles, K., Palomero-Gallagher, N., Amunts, K., 2013. Development of cortical
  folding during evolution and ontogeny. Trends in Neurosciences 36~(5),
  275--284.

\end{thebibliography}

\end{document}